%% file: main.tex
\documentclass{article}

\usepackage[final]{neurips_2021}

\usepackage[utf8]{inputenc} 
\usepackage[T1]{fontenc}    
\usepackage{hyperref}       
\usepackage{url}            
\usepackage{booktabs}       
\usepackage{amsfonts}       
\usepackage{nicefrac}       
\usepackage{microtype}      
\usepackage{amsmath}
\usepackage{float}
\usepackage{fontawesome} 
\usepackage{pifont}
\usepackage{footnote}

\makesavenoteenv{tabular}
\makesavenoteenv{table}

\usepackage{multirow}
\usepackage{graphicx}
\title{Exploring Popularity Bias in Session-based Recommendation}

\author{
  Haowen Wang\\
  Alipay AntGroup \\
  \texttt{wanghaowen.whw@antgroup.com}
}

\begin{document}
 
\maketitle

\begin{abstract}
Existing work has revealed that large-scale offline evaluation of recommender systems for user-item interactions is prone to bias caused by the deployed system itself, as a form of closed loop feedback. Many adopt the \textit{propensity} concept to analyze or mitigate this empirical issue. In this work, we extend the analysis to session-based setup and adapted propensity calculation to the unique characteristics of session-based recommendation tasks. Our experiments incorporate neural models and KNN-based models, and cover both the music and the e-commerce domain. We study the distributions of propensity and different stratification techniques on different datasets and find that propensity-related traits are actually dataset-specific. We then leverage the effect of stratification and achieve promising results compared to the original models.
%
\end{abstract}

\section{Introduction}

Recommender systems (RS) have played an important role in many successful businesses, offline testing is with no doubt an essential part of its development pipeline. Compared to online testing, a robust offline evaluation method can help make efficient model deployment decisions and save unnecessary overhead in business operations. However, some recent literature suggests that when evaluating a model on offline datasets, there can exist different strata of the data that give contradictory evaluation results\citet{simpson}. In specific, they found that model A outperforms model B on the whole dataset, but if they stratify the datasets into two strata based on certain nature of the data, there exists at least one stratum in the data space where model B can outperform model A. 

In this project, we investigate if such paradoxical phenomenon also exists in the domain of session-based recommendation. Particularly we want to find if there are specific types of situations where a deep-learning-based model can perform better than a KNN-based model. Our interest in this problem stems from a recent observation, which shows among session-based RS, KNN outperforms all other deep learning models on popular closed-loop datasets in retail and music \citep{empirical}. However, this might not be the whole story. We hypothesize it is possible that deep learning models can show more benefit on some niche in the dataset. There are of course many possible ways to stratify a dataset. In this project we focus on stratification based on the popularity of an item. That is to say, we want to show if deep learning models can perform better on unpopular items while a simpler model like KNN performs better on data with overall high popularity. 

For the rest of the paper, we will focus on two research questions. First, how a deep learning model performs compared to a KNN model on data with different popularity level. To test this, we stratify datasets into two subsets by the popularity of each item indicated by a measure called the propensity score. Then we evaluate models on these two strata and compare their performances respectively. We repeat the experiments on two retail datasets and one music dataset. The result we find is that there do exist such strata where a deep learning model can outperform a KNN model on at least one stratum and underperform on the other. Second, how can we a develop new model exploiting this performance difference? We ensemble a KNN model and a deep learning model based on the propensity score criterion for stratification and find that our ensembled model outperformed both SKNN and GRU4REC by 3-5\% on retail datasets. We believe our work is innovative and impactful in a sense that it provides a novel framework of multi-dimensional model evaluation. This framework will allow companies to identify where a model’s strengths are for better model choice and provoke thoughts in the research community for future research directions.

\section{Related Work}
\textbf{Session-based Recommender Systems.}
Conventional recommender systems take a rather static approach to modelling user preferences. They treat historical user-item interactions equally and provide one-shot recommendation. There is discrepancy between this paradigm and the real-world cases: the user preferences could develop over time, which makes it necessary to distinguish the timestamp of records. This short-term preference is embedded in the user’s most recent interactions \citep{Jannach2017SessionbasedIR,wang_gace_2024}, which often account for a small proportion of her historical interactions. To cater to this realistic need, session-based recommender systems approach this problem differently. They do not model user preferences but instead tracks the sessions and make decisions based on the seen interactions of the session, where a session is a series of user-item interactions that happen continuously. For session-based tasks/datasets, the goal is to predict a partial consecutive sequence of user actions given the previous seen actions within the same session.

\textbf{KNN-based Models.}
KNN models find the K sessions that are most similar (based on a certain metric) to the target session and generate the recommendation for this session as a function of the features and labels for these K sessions. As the unit under consideration is a session, this approach requires proper feature definitions of sessions. A typical variant is VS-KNN~\citep{DBLP:journals/corr/abs-1803-09587}, which utilizes a Inverse-Document-Frequency(IDF) weighting schema when measuring the similarity between sessions.

\textbf{Neural Models.}
Methodology-wise, KNN models are non-parametric while neural models are parametric. Most neural models for session-based recommendation utilize sequence models such as RNNs in line with the sequential nature of a session. Among them, GRU4REC~\citep{gru4rec} is a widely used model that encodes the session with GRU. NARM~\citep{narm} improves over GRU4REC in session modeling with the help of a hybrid encoder with attention mechanism. SR-GNN~\citep{srgnn} models sessions as directed graphs and employ graph neural networks as opposed to sequence models and is better at capturing the transitions between items.

\textbf{Unbiased Recommender System Evaluation.} 
In offline evaluation of recommender systems, the deployed recommender system is a confounding factor that affects both the exposure of items and the user selection. Therefore it would be hard to distinguish which user interactions stem from the users’ true preferences and which are influenced by the deployed recommender system. To remedy this bias, unbiased evaluation methods take the frequency of items in the training dataset into account and assign different weights to evaluation instances so that the final result is more correlated to open-loop evaluations. \citet{propensity1} adopt Inverse-Propensity-Scoring (IPS) technique, where the propensity of a user-item pair represents the frequency that an item is presented to a user. \citet{propensity2} relax the definition of propensity and makes it a general score that is unaware of the user of interest. This is the definition that we follow in our work.


\section{Methodology}
\subsection{Propensity Score Calculation}
In this section, we will explain in detail our methodology of calculating the propensity scores, based on which we stratified our data in the initial experiments. As stated in \citet{simpson}, a propensity score $p_{i,u}$ is defined as the probability that the deployed model exposes item i to user u under a closed loop feedback scenario. Since the true probability for a model exposing some items is always unknown to us, we will estimate the propensity score using the raw observations from the dataset with a method proposed by \citet{propensity2}. Based on their framework, we introduce the following assumptions and observations. 

\textbf{Assumption 1:} the propensity score $p_{i,u}$ is user-independent. This assumption was first made in \citet{propensity2} to address the lack of auxiliary user information in many user-item interaction records. We will keep this assumption in our method because user information is usually not utilized anyways under a session-based context. With this assumption, we are able to write the propensity score of a certain item as 
\begin{equation}
    p_{*,i} = p_{*,i}^{select} \cdot p_{*,i}^{interact|select}
\end{equation}
where $p_{*,i}^{select}$ is the probably that item $i$ is recommended and $p_{*,i}^{interact|select}$ is the conditional probability of the user interacting with item $i$ given that it is recommended.

\textbf{Assumption 2:} $p_{*,i}^{interact|select} = p_{*,i}^{interact}.$ That is, a user’s preference over items is not affected by what is recommended to her. That makes the probability that a user will interact with an item conditioned on it being recommended equal to the true probability that a user will interact with the item. Since this probability is user independent, it is proportional to the item’s true popularity $n_i$: 
\begin{equation}
    \hat{p}_{*,i}^{interact} \propto n_i
\end{equation}

\textbf{Assumption 3:} $p_{*,i}^{select}$ is proportional to $(n_i^*)^\gamma$, where $n_i^*$ is the number of observed interactions. This assumption comes from a common template of popularity bias introduced in \citet{Popularity} based on empirical observations. 

With one more observation that $n_i^*$ is sampled from a binomial distribution parameterized by $n_i$, we are able to write the estimated propensity score of an item as 
\begin{equation}
\hat{p}_{*,i} \propto (n_i^*)^{\frac{\gamma+1}{2}}
\label{eq:propensity}
\end{equation}

In the following section, we used this method to calculate the propensity score for each item in the dataset, with one caveat: instead of stratifying our dataset with one item as a basic unit, we need to calculate context-aware propensity for each action that aggregates the item propensity of all previously seen items in the current session.

\subsection{Evaluation Metrics}
We follow the evaluation metrics in \citet{empirical} to evaluate each model’s performance on stratified test data. We will focus on the following main metrics, since non-neural network models consistently outperform neural network models in terms of them.

\textbf{Hit Rate (HR)} is the percentage of actions where the model's predicted item is the target item. 

\begin{equation}
    HR=\frac{|A_{hit}^{L}|}{|A_{all}|}
\end{equation}



\textbf{Mean Reciprocal Rank (MRR)} measures the relevance of recommendations by taking reciprocal rank of each recommended item and averaging over all actions. 
$$MRR = \frac{1}{|A_{all}|}\Sigma_{u=1}^{|A_{all}|}RR(a)$$
$$RR = \Sigma_{i\leq L} \frac{relevance_{i}}{rank_i}$$

\subsection{Models}
There are mainly two models involved in our work. The SKNN represents nearest neighbor based models and GRU4REC~\cite{gru4rec} represents the deep-learning based models.
\subsubsection{SKNN}
The session based KNN method generates recommendation based on session similarity of the current session and other known sessions. The metric used to compute the similarity between sessions in our project is the Jaccard distance. The Jaccard distance between two session S1 and S2 is:
$$d_{jaccard}(S1,S2)=\dfrac{|S1 \cap S2|}{|S1 \cup S2|}$$
After the top K similar sessions in the training set are selected, the SKNN algorithm recommends items based on item popularity in this subset. \\
Various extensions of the SKNN algorithm incorporates ideas like IDF, Inverse Document Frequency. However, our exploratory results shows that these methods do not behave significantly different or always outperform SKNN, therefore we used SKNN as the representation of the KNN based model for its popularity.
\subsubsection{GRU4REC}
\begin{figure}
    \centering
    \includegraphics[width=0.8\linewidth]{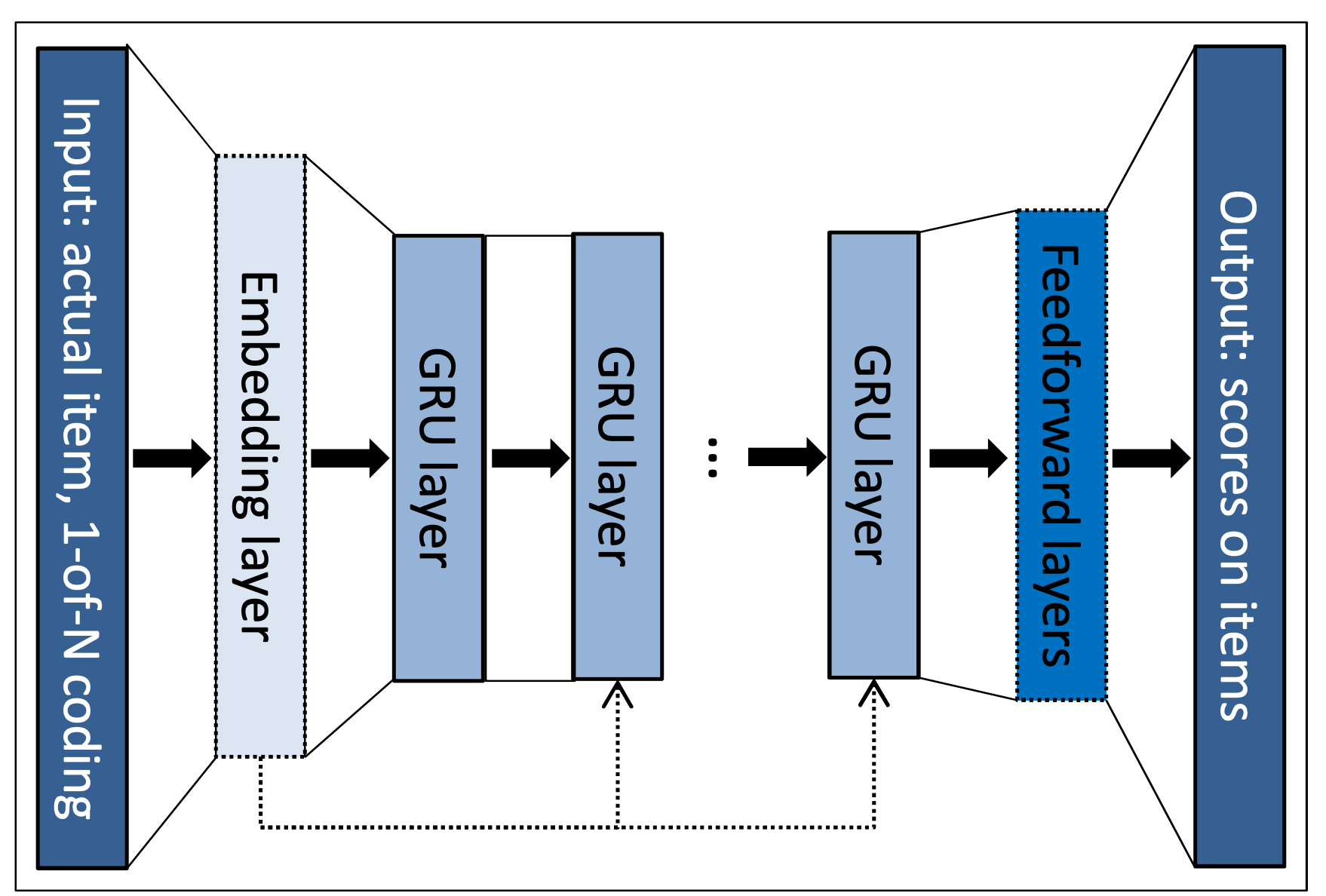}
    \caption{The architecture of the GRU4Rec network.\cite{gru4rec}.The input embedding of the model is an one-hot embedding of items}
    \label{fig:gru4rec}
\end{figure}

The GRU4REC is one of the first approach that uses Recurrent Neural Networks for session-based recommender systems and it is used widely as an strong baseline RNN model in the field. The architecture of the GRU4REC network is shown in figure \ref{fig:gru4rec}. 

The authors of GRU4REC have extended their work in 2018 \cite{gru4rec2}, where they investigated several net loss functions used and reported a 35\% performance increase using the new loss functions. In our project, we uses the Bayesian Personalized Ranking (BPR) based loss function. 
\section{Experiments}

\subsection{Datasets}
\input{ds}
For our experiments, we adopt two e-commerce recommendation datasets, Diginetica and Retailrocket. These datasets are of the same domain and moderate size. Additionally, we adopt 30MU dataset that contains the music listening log obtained from Last.fm to diversify the domain of the our experiments. The specifications are given in Table \ref{tab:ds}. Compared to the other two datasets, 30MU generally has longer sessions and more involved items. Following previous work, we remove sessions with length one, as well as items that are interacted with less than 5 times.

\subsection{Evaluation on Stratified Datasets}
\label{sec:stratify}
To answer the first question of whether such strata exist in session-based recommendation system, we leverage the evaluation method proposed in \citet{simpson}. The evaluation method calculates propensity score for each item and stratifies each \textit{action} by context-aware propensity. 

To run the approach to session-based recommendation system datasets, we design two methods to calculate score for each action in a session. And we use these two methods to split the evaluation data into two parts, Q1 (actions with low score) and Q2 (actions with high score).

\textbf{Stratification based on the propensity of the target item.} In this method, we directly use the propensity of the target (ground truth) item. But in reality, our algorithm is not supposed to have access to an action's target item until the evaluation of this action is done. So, this is an ideal method but somehow violates the evaluation protocol by peeking into the future.

\textbf{Stratification based on the propensities of historical items in the session.} In this method, instead of using the information from the target item, we use the information of all previous seen items in the session. A straightforward way to aggregate these items is to take the average of the propensity scores of the historical items in the session. Here, we make the assumption that this metric is correlated to the propensity of the target item. (In Figure~\ref{fig:corr}, we provide a visualization to demonstrate the empirical correlation between the metrics for each dataset).

\subsection{Model Training and Evaluation}
For our experiments, we used SKNN as our KNN-based models and GRU4REC as the deep-learning based model. We decided to include them in our experiments because VKNN model itself has long been favored by practitioners We used GRU4REC as the neural-net model for evaluation because as it is one of the pioneers and often served as a strong baseline for new neural network models in the field. 

Our training and inference framework were primarily inherited from the Python framework \textbf{session-rec}, we implemented additional propensity-score based ensemble models, inference scripts and rewritten parts of evaluation classes to suit our evaluation goals better. The hyper-parameters we have chosen for each model were mainly coming from \cite{empirical} and are listed here:\\
\begin{enumerate}
    \item For SKNN, we used k=100, sample size of 500 and cosine similarity. 
    \item For GRU4REC, we used the bpr-max loss function \citep{gru4rec}, dropout rate of 0.3, learning rate of 0.03 and momentum of 0.1. The learning algorithm was Adagrad.
\end{enumerate}




\section{Results and Discussion}
\subsection{Item Propensity Distribution}

\begin{figure}[h]
	\centering
	\includegraphics[width=1\linewidth]{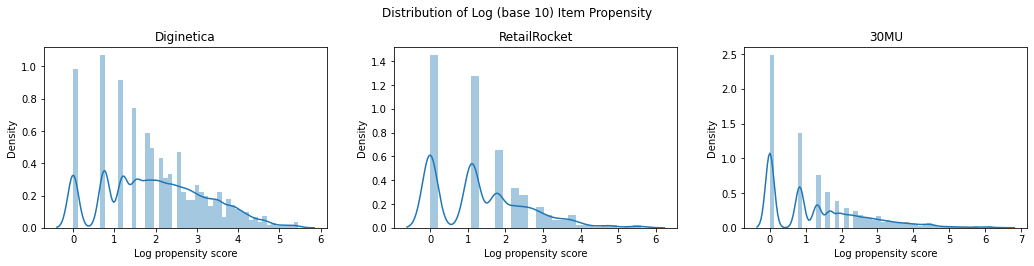}
	\caption{Visualization of the distribution of item propensity of three datasets, in log scale (base 10).}
	\label{fig:item_propensity}
\end{figure}
We first visualize the item propensity for each dataset to grasp their characteristics. The calculation of item propensity is based on Eq.~\ref{eq:propensity} and the power law algorithm is fitted independently for each dataset.

We show the distribution of log item propensity for the three datasets in Fig.~\ref{fig:item_propensity}. Since we use the power law, the range of the resultant item propensity is similar in different datasets. However, the distribution of Diginetica and RetailRocket is more spread out and that of 30MU is more centered toward the lower end. 

\subsection{Propensity-Based Evaluation}
\begin{figure}[h]
	\centering
	\includegraphics[width=0.85\linewidth]{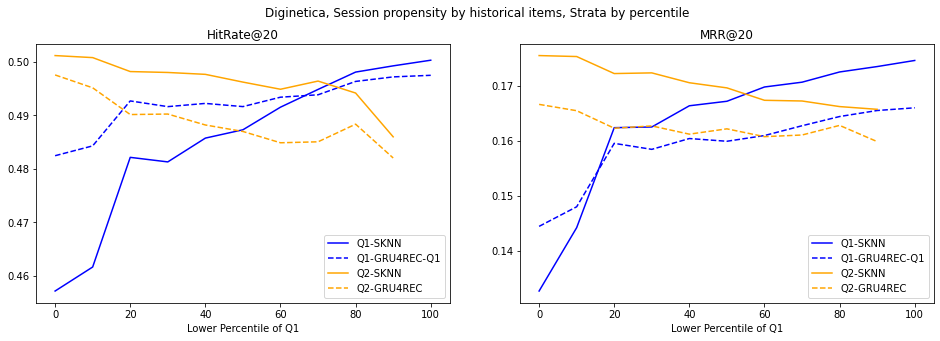}
	\caption{Different models evalueted on Diginetica strata generated by different percentile cutoff. The average of historical item propensity as action-wise propensity.}
	\label{fig:dig_historical}
\end{figure}
\begin{figure}[h]
	\centering
	\includegraphics[width=0.85\linewidth]{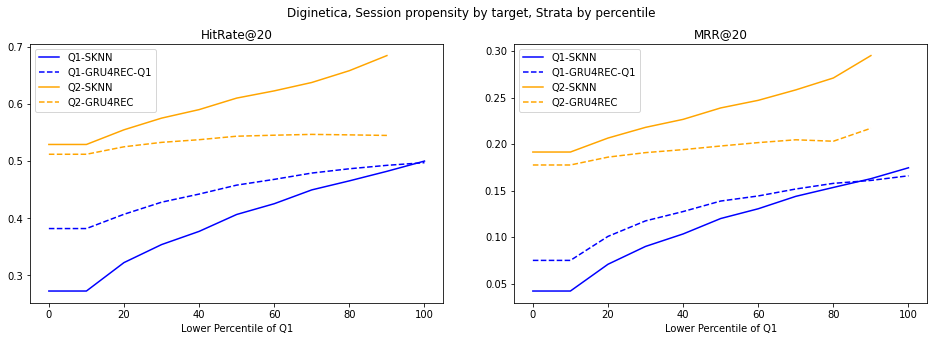}
	\caption{Different models evaluated on Diginetica strata generated by different percentile cutoff. The item propensity of target item as action-wise propensity.}
	\label{fig:dig_target}
\end{figure}

We stratify evaluation data based on action-wise propensity into varying proportions and report the results of different models on different strata. 

Specifically, we calculate the action-wise propensity score using the two methods elaborated in Sec.~\ref{sec:stratify}. And we experiment with different $x$ values such that the $x\%$ percentile of propensity score value split the whole evaluation dataset into two portions, Q1 and Q2. We then evaluate two different models, SKNN and GRU4REC on both portions to analyze the results of different models on actions with different propensities.

The results for Dignetica are shown in Fig.~\ref{fig:dig_historical} and Fig.~\ref{fig:dig_target}. Due to space limit, we move the results for RetailRocket and 30MU to the Appendix (Sec~\ref{sec:additional_propensity_based}). For e-commerce datasets (Dignetica and RetailRocket), as the percentile decreases, which means the average popularity of items decreases, GRU4REC clearly gains advantage over SKNN. However, in 30MU, there is no such trend and the gap between the two models is consistent across different $x$.

\begin{figure}
	\centering
	\includegraphics[width=1\linewidth]{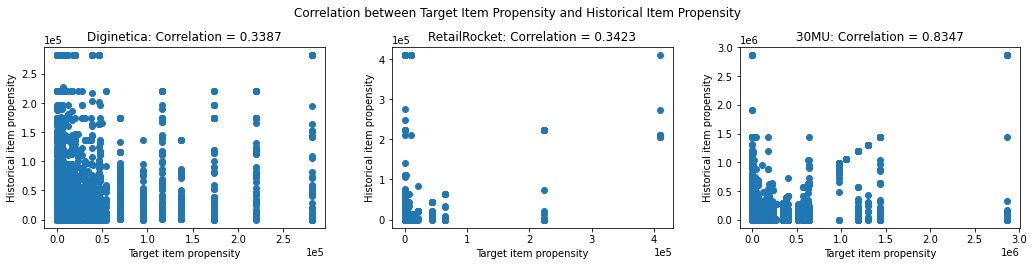}
	\caption{Correlation between target item propensity and the average of historical item propensities.}
	\label{fig:corr}
\end{figure}

Additionally, we visualize the correlation between the two stratification methods, shown in Fig.~\ref{fig:corr}. There is strong correlation in 30MU dataset and low correlation in the two e-commerce datasets. However, the correlation is not necessarily related to the experimental results we achieve- even for 30MU, the trends from the line plot in Fig.~\ref{fig:music_historical} \& \ref{fig:music_target} are not alike for these two stratification methods.

\input{tab/ratio}
To examine the robustness of deep learning models against unpopular items, we take the S1, the set of actions with bottom 10\% propensity score, and S2, the set of actions with top 10\% propensity score. We evaluate the performance of the two models on each set, and compute the ratio between score of S1 and score of S2. The result is shown in Table~\ref{tab:ratio}. GRU4REC has higher ratio compared to SKNN for all datasets, suggesting that it is more robust against unpopular items.

\section{Model Ensemble}
Since there is a general trend of GRU4REC outperforming SKNN for actions with lower propensity score and SKNN outperforming GRU4REC for actions with higher propensity score, we experiment with model ensemble based on the action-wise propensity in evaluation. We adopt the stratification method based on the item propensities of historical items in the session (Sec.~\ref{sec:stratify}).

\subsection{Fixed-Weight Ensemble Methods}
Specifically, for each action, we set a threshold based on our analysis of the action-wise propensity score distribution, which is dataset-specific. We assign different weights to the prediction scores of GRU4REC and SKNN model and use the weighted average as the final prediction score of the action. The weights and the threshold are hyper-parameters. We use a symmetrical weight assignment approach where the weights for GRU4REC and SKNN are reversed in two strata.

\input{tab/by_strata}

The ensemble method clearly benefits e-commerce datasets, bringing $~10\%$ relative improvement over HitRate@20 on Dignetica and $~6\%$ over HitRate@20 on RetailRocket. And the gain is not sensitive to the ensemble weights. On the 30MU dataset, although the GRU4REC model outperforms SKNN on every strata, the ensemble methods still benefit from incorporating SKNN's predictions.
\subsection{Dynamic-Weight Ensemble Methods}
In the previous method, we fix the weight of different models on different propensity strata, where all the actions in the same strata, regardless of its real propensity score, will be assigned the same set of weights for ensembles. Therefore, we propose the dynamic-weight ensemble method where the weights of each action are dynamically computed using their propensity score. This ensures each action gets a set of weights that better reflects its propensity. The weights $w_1, w_2$ are computed with the following formula, denote the propensity score of action $i$ as $p_i$:
\begin{equation}
    p_{normalized,i}=\dfrac{log(p_i+10^{-9})-\hat{p}_{mean}}{\hat{p}_{std}}
\end{equation}
\begin{equation}
    w_1 = \dfrac{1}{1+e^{p_{normalized, i}+\alpha}}, w_2=1-w_1
\end{equation}
The formula is essentially an x-shifted Sigmoid transformation on the negation of the log-normalized propensity score. The hyper-parameters are $\alpha, \hat{p}_{mean}$ and $\hat{p}_{std}$. $\hat{p}_{mean}$ and $\hat{p}_{std}$ are the approximation of the mean and standard deviation of the actions in the entire test dataset. It can be approximated using the actions in the training set, since the items in two sets are completely overlapped. The $\alpha$ governs how much does the Sigmoid function shifts in the x direction, it is optimized using grid search.\\
The results of our dynamic-weight ensemble methods is shown in table \ref{tab:dynamic_weight}. It is compared with three other fixed-weight ensemble methods. We could see that the dynamic-weight method gives the highest HitRate among all the methods but the MRR metric did not improve as much.

\section{Conclusion}
In this work, we test the effect of popularity bias on model performance by reevaluating KNN-based and deep-learning-based-models on the datasets stratified by propensity score. By experimenting with different stratification thresholds, we find that on the two ecommerce datasets used, GRU4REC does outperform SKNN on less popular items. SKNN, on the other hand, still outperforms GRU4REC on more popular items.  On the music dataset however, GRU4REC consistently outperforms SKNN in all strata of the dataset. Both observations align with our original hypothesis that GRU4REC can show robustness against popularity bias because it is achieving good performance on the unpopular items than KNN. We also notice that the correlation between the average of historical item propensities and the propensity of target item is much more correlated than what we observe from the two ecommerce datasets. We also take advantage of the performance discrepancy of the two models on different strata by ensembling them with fixed and dynamic weights. What we find is that ensembling the models with fixed weighting based on propensity score can improve the overall model performance and dynamic weighting further strengthens this effect. We hope our work can provoke readers to rethink about closed-loop evaluation in session-based recommendation and performance difference among each type of session-based algorithm. We also hope our work provides readers with a fresh angle of looking at popularity bias in RS datasets. Though most of the previous works discuss how we should be aware of this type of bias and avoid it when necessary, we instead incorporate it as a useful strategy for model innovation.

\bibliography{neurips}
\bibliographystyle{natbib}

\appendix

\subsection{Propensity-Based Evaluation on RetailRocket and 30MU}
\label{sec:additional_propensity_based}
\begin{figure}[h]
	\centering
	\includegraphics[width=0.85\linewidth]{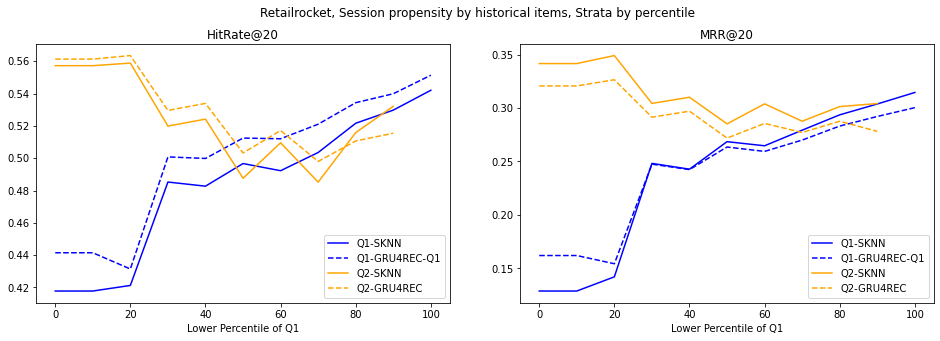}
	\caption{Different models evalueted on RetailRocket strata generated by different percentile cutoff. The average of historical item propensity as action-wise propensity.}
	\label{fig:rr_historical}
\end{figure}
\begin{figure}[h]
	\centering
	\includegraphics[width=0.85\linewidth]{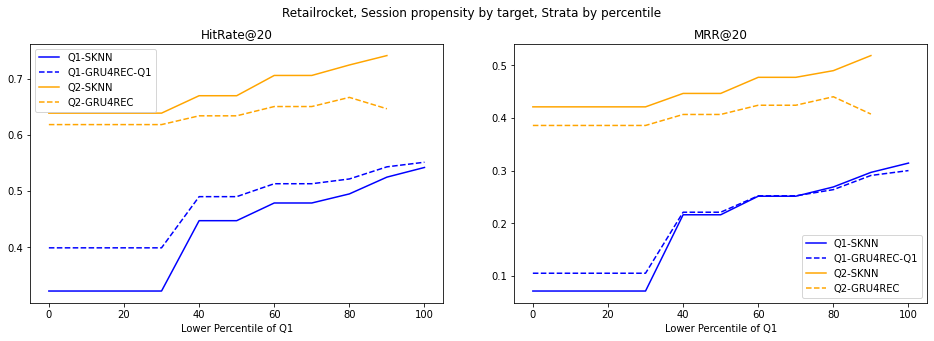}
	\caption{Different models evaluated on RetailRocket strata generated by different percentile cutoff. The item propensity of target item as action-wise propensity.}
	\label{fig:rr_target}
\end{figure}

\begin{figure}[h]
	\centering
	\includegraphics[width=0.85\linewidth]{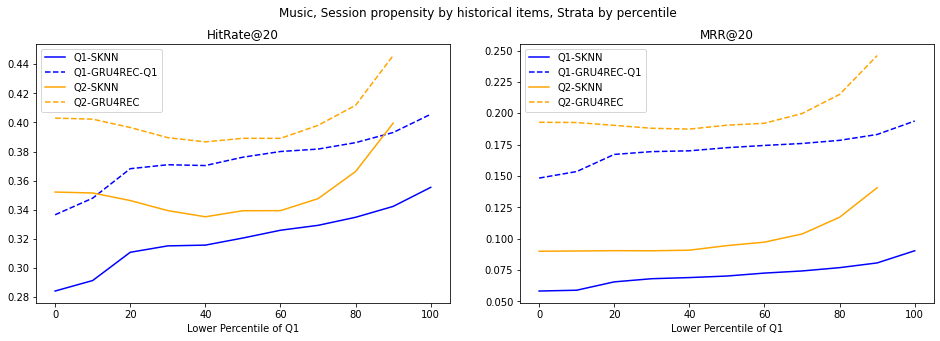}
	\caption{Different models evalueted on 30MU strata generated by different percentile cutoff. The average of historical item propensity as action-wise propensity.}
	\label{fig:music_historical}
\end{figure}
\begin{figure}[h]
	\centering
	\includegraphics[width=0.85\linewidth]{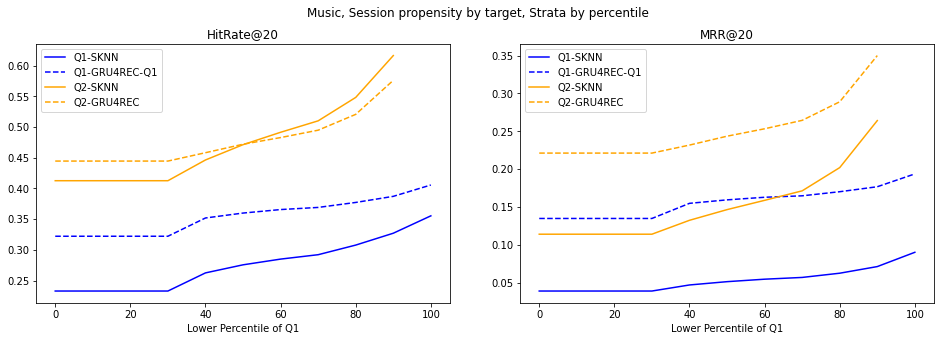}
	\caption{Different models evaluated on 30MU strata generated by different percentile cutoff. The item propensity of target item as action-wise propensity.}
	\label{fig:music_target}
\end{figure}



\end{document}

%% file: ds.tex
\begin{table*}[h]
\centering
\begin{tabular}{lccc}
    \toprule  
    \textbf{Dataset} & Diginetica & Retailrocket & 30MU \\
    \midrule  
    Actions  &  264k  &  210k & 640k \\
    Sessions  &  55k  &  60k & 37k \\
    Items &  32k & 32k & 91k \\
    \midrule
    Avg. Session Length & 4.78 & 3.54 & 17.11\\
    \bottomrule 
\end{tabular}
\vspace{-0.1cm}
\caption{Data specifications of datasets.}
\label{tab:ds}
\end{table*}

%% file: tab/ratio.tex
\begin{table*}[ht]
\centering
\scalebox{0.9}{
\begin{tabular}{lccc}
    \toprule  
    \textbf{Dataset} & \textbf{Metric} & \textbf{SKNN} & \textbf{GRU4REC}  \\
    \midrule  
    \multirow{2}{*}{\textit{DIGINETICA}} & HitRate@20 & 0.3978 & 0.7007  \\
    \cmidrule(lr){2-4} & MRR@20 & 0.1424 & 0.3455  \\
    \midrule
    \multirow{2}{*}{\textit{RetailRocket}} & HitRate@20 & 0.4351 & 0.6175 \\
    \cmidrule(lr){2-4} & MRR@20 & 0.1383 & 0.2589  \\
    \midrule
    \multirow{2}{*}{\textit{30MU}} & HitRate@20 & 0.7293 &0.7803  \\
    \cmidrule(lr){2-4} & MRR@20 & 0.4187 & 0.6241   \\
    
    \bottomrule 
\end{tabular}
}
\caption{We report the ratio between model performance on S1, the set of actions with bottom 10\% propensity score, and model performance on S2, the set of actions with top 10\% propensity score. For each dataset and for each metric, GRU4REC shows a higher bottom-to-top ratio. This result shows the robustness of GRU4REC on less popular items, across all datasets we examine.}
\label{tab:ratio}
\end{table*}

%% file: tab/by_strata.tex
\begin{table*}[ht]
\centering
\scalebox{0.9}{
\begin{tabular}{lccccccccc}
    \toprule  
    \multirow{2}{*}{\textbf{Dataset}} & \multirow{2}{*}{\textbf{Metric}} & \multirow{2}{*}{\textbf{SKNN}} & \multirow{2}{*}{\textbf{GRU4REC}} & \multicolumn{6}{c}{\textbf{Ensemble, weight} $w_2$}  \\ \cmidrule(lr){5-10} & & & & $1.0$ & $0.9$ & $0.8$ & $0.7$ & $0.5$ & $0.2$ \\
    \midrule  
    \multirow{2}{*}{\textit{DIGINETICA}} & HitRate@20 & 0.500 & 0.497 & 0.504 & 0.548 & \textbf{0.553} & 0.550 & 0.540 & 0.520 \\
    \cmidrule(lr){2-10} & MRR@20 & 0.175 & 0.166 & 0.175 & 0.183 & 0.186 & 0.187 & \textbf{0.188} & 0.180 \\
    \midrule
    \multirow{2}{*}{\textit{RetailRocket}} & HitRate@20 & 0.548 & 0.551 & 0.550 & \textbf{0.581} & \textbf{0.581} & 0.578 & 0.577 & 0.565 \\
    \cmidrule(lr){2-10} & MRR@20 & 0.311 & 0.300 & 0.320 & \textbf{0.326} & \textbf{0.326} & 0.325 & 0.323 & 0.310 \\
    \midrule
    \multirow{2}{*}{\textit{30MU}} & HitRate@20 & 0.355 &0.4057 & - & - & \textbf{0.420} & - & - & - \\
    \cmidrule(lr){2-10} & MRR@20 & 0.09 & \textbf{0.194} & - & - & 0.160 & - & - & - \\
    
    \bottomrule 
\end{tabular}
}
\vspace{-0.1cm}
\caption{We report the evaluation results in the model ensemble scenario. The 10\% percentile threshold $T$ is $12.425$ for Dignetica, $1.000$ for RetailRocket, $110.085$ for 30MU. $w_2$ is the weight of the GRU4REC model prediction for actions with low propensity (and the weight of SKNN model prediction for actions with high propensity). We also report the result for single model evaluation for comparison.}
\label{tab:by_strata}
\end{table*}

\begin{table*}[ht]
\centering
\scalebox{0.9}{
\begin{tabular}{lccccc}
    \toprule  
    \multirow{2}{*}{\textbf{Dataset}} & \multirow{2}{*}{\textbf{Metric}} & \multirow{2}{*}{\textbf{Ensemble, dynamic weight}} & \multicolumn{3}{c}{\textbf{Ensemble, fixed weight} $w_2$}  \\ \cmidrule(lr){4-6} & & & $1.0$ & $0.8$ &$0.5$ \\
    \midrule  
    \multirow{2}{*}{\textit{DIGINETICA}} & HitRate@20 & \textbf{0.556} & 0.504  & 0.553 & 0.520 \\
    \cmidrule(lr){2-6} & MRR@20 & \textbf{0.186} & 0.175 &  \textbf{0.186} & 0.180 \\
    \midrule
    \multirow{2}{*}{\textit{RetailRocket}} & HitRate@20 & \textbf{0.586} & 0.550 & 0.581  & 0.565 \\
    \cmidrule(lr){2-6} & MRR@20 & 0.324 & 0.320 &  \textbf{0.326} &  0.310 \\
    \bottomrule 
\end{tabular}
}
\vspace{-0.1cm}
\caption{We report the result comparison between the fixed weight ensemble method versus the dynamic weight ensemble method.}
\label{tab:dynamic_weight}
\end{table*}

%% file: main.bbl
\begin{thebibliography}{12}
\expandafter\ifx\csname natexlab\endcsname\relax\def\natexlab#1{#1}\fi

\bibitem[{Hidasi and Karatzoglou(2018)}]{gru4rec2}
Bal{\'a}zs Hidasi and Alexandros Karatzoglou. 2018.
\newblock Recurrent neural networks with top-k gains for session-based
  recommendations.
\newblock In \emph{Proceedings of the 27th ACM international conference on
  information and knowledge management}, pages 843--852.

\bibitem[{Hidasi et~al.(2016)Hidasi, Karatzoglou, Baltrunas, and
  Tikk}]{gru4rec}
Balázs Hidasi, Alexandros Karatzoglou, Linas Baltrunas, and Domonkos Tikk.
  2016.
\newblock \href {http://arxiv.org/abs/1511.06939} {Session-based
  recommendations with recurrent neural networks}.

\bibitem[{Jadidinejad et~al.(2021)Jadidinejad, Macdonald, and Ounis}]{simpson}
Amir~H. Jadidinejad, Craig Macdonald, and Iadh Ounis. 2021.
\newblock \href {http://arxiv.org/abs/2104.08912} {The simpson's paradox in the
  offline evaluation of recommendation systems}.

\bibitem[{Jannach et~al.(2017)Jannach, Ludewig, and
  Lerche}]{Jannach2017SessionbasedIR}
D.~Jannach, Malte Ludewig, and Lukas Lerche. 2017.
\newblock Session-based item recommendation in e-commerce: on short-term
  intents, reminders, trends and discounts.
\newblock \emph{User Modeling and User-Adapted Interaction}, 27:351--392.

\bibitem[{Li et~al.(2017)Li, Ren, Chen, Ren, and Ma}]{narm}
Jing Li, Pengjie Ren, Zhumin Chen, Zhaochun Ren, and Jun Ma. 2017.
\newblock \href {http://arxiv.org/abs/1711.04725} {Neural attentive
  session-based recommendation}.

\bibitem[{Ludewig and Jannach(2018)}]{DBLP:journals/corr/abs-1803-09587}
Malte Ludewig and Dietmar Jannach. 2018.
\newblock \href {http://arxiv.org/abs/1803.09587} {Evaluation of session-based
  recommendation algorithms}.
\newblock \emph{CoRR}, abs/1803.09587.

\bibitem[{Ludewig et~al.(2020)Ludewig, Mauro, Latifi, and Jannach}]{empirical}
Malte Ludewig, Noemi Mauro, Sara Latifi, and Dietmar Jannach. 2020.
\newblock \href {https://doi.org/10.1007/s11257-020-09277-1} {Empirical
  analysis of session-based recommendation algorithms}.
\newblock \emph{User Modeling and User-Adapted Interaction}, 31(1):149–181.

\bibitem[{Schnabel et~al.(2016)Schnabel, Swaminathan, Singh, Chandak, and
  Joachims}]{propensity1}
Tobias Schnabel, Adith Swaminathan, Ashudeep Singh, Navin Chandak, and Thorsten
  Joachims. 2016.
\newblock \href {http://arxiv.org/abs/1602.05352} {Recommendations as
  treatments: Debiasing learning and evaluation}.

\bibitem[{Steck(2011)}]{Popularity}
Harald Steck. 2011.
\newblock \href {https://doi.org/10.1145/2043932.2043957} {Item popularity and
  recommendation accuracy}.
\newblock In \emph{Proceedings of the Fifth ACM Conference on Recommender
  Systems}, RecSys '11, page 125–132, New York, NY, USA. Association for
  Computing Machinery.

\bibitem[{Wang et~al.(2024)Wang, Du, Jin, Li, Wang, Sun, Qin, and
  Fan}]{wang_gace_2024}
Haowen Wang, Yuliang Du, Congyun Jin, Yujiao Li, Yingbo Wang, Tao Sun, Piqi
  Qin, and Cong Fan. 2024.
\newblock {GACE}: {Learning} {Graph}-{Based} {Cross}-{Page} {Ads} {Embedding}
  for {Click}-{Through} {Rate} {Prediction}.
\newblock In \emph{Neural {Information} {Processing}}, pages 429--443,
  Singapore. Springer Nature Singapore.

\bibitem[{Wu et~al.(2019)Wu, Tang, Zhu, Wang, Xie, and Tan}]{srgnn}
Shu Wu, Yuyuan Tang, Yanqiao Zhu, Liang Wang, Xing Xie, and Tieniu Tan. 2019.
\newblock \href {https://doi.org/10.1609/aaai.v33i01.3301346} {Session-based
  recommendation with graph neural networks}.
\newblock \emph{Proceedings of the AAAI Conference on Artificial Intelligence},
  33:346–353.

\bibitem[{Yang et~al.(2018)Yang, Cui, Xuan, Wang, Belongie, and
  Estrin}]{propensity2}
Longqi Yang, Yin Cui, Yuan Xuan, Chenyang Wang, Serge~J. Belongie, and Deborah
  Estrin. 2018.
\newblock Unbiased offline recommender evaluation for missing-not-at-random
  implicit feedback.
\newblock \emph{Proceedings of the 12th ACM Conference on Recommender Systems}.

\end{thebibliography}
